\documentstyle[eqsecnum,aps]{revtex}
\begin{document}
\title{On the Evidence for Postacceleration Effects\\
in Breakup Reactions with Halo Nuclei}
\author{C.A. Bertulani$^{a)}$, M.S. Hussein$^{b)}$, M.P. Pato$^{b)}$, and M. Ueda$%
^{b)}$}
\address{$^{a)}$ Instituto de F{\'\i}sica, Universidade Federal do Rio de Janeiro,
Caixa Postal 68528, \\
21945-970, Rio de Janeiro, Brazil\\
$^{b)}$ Instituto de F{\'\i}sica, Universidade de S{\~a}o Paulo, Caixa Postal 66318,\\
05389-970 S{\~a}o Paulo SP, Brazil}
\date{\today }
\maketitle

\begin{abstract}
We study the postacceleration of charged fragments in reactions with
unstable nuclear beams. For elastic breakup processes we show that the
postaceleration effect can be well understood in terms of closed analytical
forms derived in a quantum mechanical formulation. This gives theoretical
support to the effect experimentally observed in the breakup of $^{11}Li$
projectiles at intermediate energies.
\end{abstract}

\bigskip


\bigskip

%\twocolumn

Over the last four years an extensive effort was made, both experimentally
and theoretically, to understand the break-up mechanism of halo nuclei
[1-9]. In particular, Refs.[1,2], reported a measurement of the parallel
velocity distribution of $^9$Li in the elastic break-up of $^{11}$Li at 28
MeV$\cdot $A, as predicted in ref. \cite{3a} by means of semiclassical
arguments. This distribution showed an asymmetry with respect to the
parallel velocity distributions of the neutrons and of the $^9Li$ fragments: 
$^9$Li comes out faster than the neutrons as it splits from $^{11}$Li. This
was interpreted as a postacceleration due to the Coulomb interaction in the
final state. This observation was used to deduce the nature of the so-called
soft dipole mode. The authors of Refs.[1,2] came to the conclusion that this
mode of excitation is not a resonance since the break-up seems to have
occurred in the vicinity of the charged field of the target. However, such
an interpretation came under questioning in a recent publication that
discussed the break-up and postacceleration of the also `` exotic'' deuteron 
\cite{7}, and also a DWBA \cite{8} and a semiclassical \cite{9} calculation.

It is clear that more data are required in order to better understand the
postacceleration phenomenon. Simple closed form models that explicitly
exhibit the effect are certainly welcome as they will supply a guide to
experimentalists. The knowledge of the width of the parallel velocity
distribution is particularly important since it allows the preparation of
experimental setups. The purpose of this note is to present such a simple
description of the postacceleration effect.

Our starting point is the prior form of the Distorted Wave Born
Approximation ( DWBA ) description of elastic break-up of nuclear
projectile. The amplitude is given by, for the $^{11}$Li $\longrightarrow $ $%
^9$Li $+2$n reaction, 
\begin{equation}
T_{if}=<\chi _2^{(-)}({\bf r_2})\chi _9^{(-)}({\bf r_9})\left| \left[ U_2(%
{\bf r_2})+U_9({\bf r_9})-U_{11}({\bf R})\right] \right| \chi _{11}^{(+)}(%
{\bf R})\varphi _{g.s.}({\bf r})>
\end{equation}
where $U_i({\bf r_i})$ is the complex optical potential of nucleus $i$, and $%
\chi _i^{(\pm )}({\bf r_i})$ is the corresponding optical wave function with
outgoing $(+)$ and incoming $(-)$ wave boundary conditions. At intermediate
energies ($\sim 50$ MeV/nucleon), the Sommerfeld parameter $\eta
=Z_PZ_Te^2/\hbar v\gg 1$, and the reaction is very forward peaked, so that
we can employ the eikonal form for these wave functions, viz., 
\begin{eqnarray}
\chi _{11}^{(+)}({\bf R}) &=&e^{i{\bf k_{11}}\cdot {\bf R}}\exp {\left[ -%
\frac i{\hbar v_{11}}\int_{-\infty }^{z_{11}}U_{11}({\bf b_{11}}%
,z_{11}^{\prime })dz_{11}^{\prime }+i\phi _{11}^{(c)}(b_{11},\
z_{11})\right] }  \nonumber \\
&& \\
\chi _9^{(-)*}({\bf r_9}) &=&e^{-i{\bf k_9}\cdot {\bf r_9}}\exp {\left[ -%
\frac i{\hbar v_9}\int_{z_9}^\infty U_9({\bf b_9},z_9^{\prime })dz_9^{\prime
}+i\phi _9^{(c)}(b_9,\ z_9)\right] } \\
\chi _2^{(-)*}({\bf r_2}) &=&e^{-i{\bf k_2}\cdot {\bf r_2}}\exp {\left[ -%
\frac i{\hbar v_2}\int_{z_2}^\infty U_2({\bf b_2},z_2^{\prime })dz_2^{\prime
}\right] }\ ,
\end{eqnarray}
where 
\begin{equation}
\phi _{11}^{(c)}=\eta _{11}\ln \left( k_{11}(r_{11}-z_{11})\right) \ ;\ \
\phi _9^{(c)}=\eta _9\ln \left( k_9(r_9+z_9)\right) \ ;\ \ \;\;\;\;{\rm %
with\;\;\;\;}\eta _i={\frac{Z_iZ_Te^2}{\hbar v_i}}\ ,
\end{equation}
are the Coulomb phases for $i\equiv ^{11}Li,\ ^9Li$, respectively.

We now assume that the potentials in the exponents in Eqs.(2-4) are related $%
U_{11}^{(N)} = U_{9}^{(N)} + U_{2}^{(N)} $ and ignore the difference between
the quantities $1/{\hbar v_i}$. The effect of this approximation is very
small for the calculation of postacceleration, since it modifies only
slightly the nuclear phases. As we show next the post-acceleration effect
arises from the differences between the {\it Coulomb phases} for the $^9Li$
and the two neutrons.

For the interaction $\Delta V\equiv U_2+U_9-U_{11}\simeq
U_2^{(C)}+U_9^{(C)}-U_{11}^{(C)}$, we get 
\begin{equation}
\Delta V_C=Z_PZ_Te^2[\frac{R-r_9}{Rr_9}]\;.
\end{equation}
The vectors ${\bf R}\equiv {\bf r}_{11}$, ${\bf r_9}$ and ${\bf r_2}$ are
related through 
\begin{equation}
{\bf r_2}={\bf R}+\frac 9{11}{\bf r}\ \ ,\ \ \ \ {\bf r_9}={\bf R}-\frac 2{11%
}{\bf r}
\end{equation}
where ${\bf r}$ is the relative distance between the di-neutron and $^9$Li.
With Eq. (7), $\Delta V_C$ assumes the form, in the dipole approximation, 
\begin{equation}
\Delta V_C\equiv \frac 2{11}\frac{Z_PZ_Te^2}{R^2}\sqrt{\frac{4\pi }3}\ r%
\mbox{Y}_{10}({\hat{{\bf r}}})
\end{equation}

Using the results above, the DWBA amplitude can be rewritten as \cite{10} 
\begin{equation}
T_{fi}({\bf Q},{\bf q})\equiv T_{exc}({\bf q})T_{el}({\bf Q})
\end{equation}
where 
\begin{equation}
T_{exc}({\bf q})=\frac 2{11}\sqrt{\frac{4\pi }3}\ \int d{\bf r}\ e^{-i{\bf q}%
\cdot {\bf r}+\Delta \phi _c}\ r\ \mbox{Y}_{10}({\hat{{\bf r}}})\ \varphi
_{g.s.}({\bf r})
\end{equation}
and 
\begin{equation}
T_{el}({\bf Q})=Z_PZ_Te^2\int d{\bf R}\ \ e^{-i{\bf Q}\cdot {\bf R}}\ S_{11}(%
{\bf b}_{11})\ \frac 1{R^2}
\end{equation}
where 
\begin{equation}
{\bf q}\equiv \frac 9{11}{\bf k_2}-\frac 2{11}{\bf k_9\;,\;\;\;\;\;\;Q}%
\equiv {\bf k_9}+{\bf k_2}-{\bf k_{11}}
\end{equation}
and $S_{11}({\bf b})$ is the elastic S-matrix element of the projectile.

The term $\Delta \phi _c$ in the exponential appearing in eq. (10) is the
responsible for the post-acceleration effect. It is a consequence of the
independent propagation of the neutrons and of the $^9Li$ wavefunctions in
the final channel (we neglect final state interactions between the two
neutrons and the $^9Li$ fragment). This term is identically zero if one
considers the excitation to bound states (which are absent in the case of $%
^{11}Li$ projectiles), or for the breakup of the projectile into fragments
with same charge-to-mass ratios (to first order). The term $\Delta \phi _c$
is what is left, when one incorporates the Coulomb phase for the elastic
scattering, $2i\eta _{11}\ln (k_{11}b_{11})$, in the S-matrix element, $%
S_{11}({\bf b}_{11})$, by using the relations (7). Specifically, it is given
by 
\begin{equation}
\Delta \phi _c=2i\eta _{11}\ln \left( k_{11}b_{11}\right) -i\eta _{11}\ln
\left[ k_9(r_{11}-z_{11})\right] -i\eta _9\ln \left[ k_9(r_9+z_9)\right]
\simeq {\frac{2\eta _{11}(\rho +z)}{11b_{11}}}\ ,
\end{equation}
where $\rho $ is the transverse coordinate of $r$. The approximation is
valid for high energy scattering ($z_{11}\ll b_{11}$), and to first order in 
$\rho /b_{11}$ and $z/b_{11}$.

The factorization in the form of eq. (9) is only useful if we can separate
the variables ${\bf r}$ and ${\bf R}\equiv {\bf r}_{11}$ completely. This
can be done by noticing that the $1/R^2$ term in the integrand of eq. (11)
favors small values of ${\bf R}$, as usual for Coulomb excitation processes.
This is the classical equivalent of grazing collisions. Thus we set $%
b_{11}\simeq b_{min} \simeq 1.2 (A_T+A_P)^{1/3}$ fm in eq. (13).

Using a Yukawa-type function ground state wavefunction, 
\begin{equation}
\phi _{g.s.}({\bf r})=\sqrt{\frac \kappa {2\pi }}\ {\frac{e^{-\kappa r}}r}%
,\qquad \kappa ^2={\frac{2\mu E_B}{\hbar ^2}},\qquad \mu ={\frac{18}{11}}%
m_N\ ,
\end{equation}
where $E_B$ is the binding energy and $m_N$ is the nucleon mass, the
integral in eq. (10), with the approximation (13), can be done analytically,
yielding 
\begin{equation}
T_{exc}({\bf q})={\frac{16\pi }{11}}\ \sqrt{\frac{2\kappa }3}\ i\ {\frac{%
q^{\prime }{}_z^2}{(\kappa ^2+q^{\prime }{}^2)^2}}\ ,
\end{equation}
where 
\begin{equation}
{\bf q^{\prime }}={\bf q}+{\frac 2{11}}\ {\frac{\eta _{11}}{b_{min}}}\ ({%
\hat{{\bf z}}+\hat{{\bf b}}})\ ,
\end{equation}
and $\hat{{\bf z}}$ ($\hat{{\bf b}}$) is the unit vector along the
longitudinal (perpendicular) direction.

The elastic break-up cross-section is given by 
\begin{equation}
d^6\sigma =\frac{2\pi }{\hbar v_{11}}|T_{fi}|^2\ \frac{d^3qd^3Q}{(2\pi )^6}\
\delta (E_i-E_f)\ .
\end{equation}
Since we are interested in the momentum differences between the two neutrons
and the $^9Li$ fragments, we integrate over the elastic scattering momentum $%
{\bf Q}$, i.e., 
\begin{equation}
(2\pi )^3\frac{d^3\sigma }{d^3q}=\frac{2\pi }{\hbar v_{11}}\int \frac{d^3Q}{%
(2\pi )^3}\left| T_{exc}({\bf q})\right| ^2\ \left| T_{el}({\bf Q})\right|
^2\ \delta (E_i-E_f)\ ,
\end{equation}
and we make use of the completeness of the plane waves $e^{-i{\bf Q}\cdot 
{\bf R}}$. We find, 
\begin{equation}
\frac{d^2\sigma }{dq_zdq_t}={\frac{2\pi }{\hbar v_{11}}}\ C_{el}\left|
T_{exc}({\bf Q})\right| ^2
\end{equation}
where 
\begin{equation}
C_{el}=(Z_PZ_Te^2)^2\int d{\bf R}\ \left| S_{11}({\bf b}_{11})\right| ^2\ 
\frac 1{R^4}\ .
\end{equation}
Since we are only interested in the relative velocity of the fragments,
which is implicit in the momentum ${\bf q}$, we can rewrite eq. (19) as 
\begin{equation}
\frac{d^2\sigma }{dq_zdq_t}={\cal C}{\frac{q'^2_z }{(\kappa^2+
q'^2_z+q'^2_t)^4}}\ ,
\end{equation}
where ${\cal C}$ includes all constants and factors which do not depend on $%
{\bf q}$.

From eq. (21) one can immediately see where the postacceleration effect
resides. It shows that the  longitudinal momentum distribution  is shifted
by $\Delta q_z^{(0)}=2\eta _{11}/(11b_{mim})$. For $^{11}Li$ projectiles
incident on lead at 30 MeV/nucleon $\Delta q_z\simeq \left( 8\,fm\right)
^{-1}$. Using the definition of ${\bf q}$ from eq. (12), we get ${\rm v}_9-%
{\rm v}_2\simeq 0.08\,c$ which is not small compared to the beam velocity, $%
{\rm v}_{11}\simeq 0.25c$. This gives a clear theoretical explanation of the
post-acceleration effect which was verified experimentally \cite{1,2}. We
can also give a classical interpretation of this result: the
postacceleration originates from the extra-momentum gained by the $^9Li$
after the breakup, assumed to occur at the distance of closest approach.
This extra-momentum is roughly given by $\Delta p=F\ .\ \Delta t\simeq
(Z_9Z_Te^2/b_{min}^2)\ .\ (b_{min}/2{\rm v})\simeq \hbar \eta /2b_{min}$.
This also is consistent with the findings of ref. \cite{3a}, where a
semiclassical approach to this problem was undertaken. It is however, in
contrast with ref. \cite{8} where a DWBA calculation was also performed. The
approximations used there were probably the reason for the negative result.

The width of the momentum distribution is given approximately by $\Gamma =%
\sqrt{\kappa ^2+\left( \Delta q_z\right) ^2}/2$. The first term inside the
square root is due to the Fermi momentum of the ground state wavefunction.
In figure 1 we plot $\Gamma $ and $\Delta q_z$ as a function of the
bombarding energy of $^{11}Li$ projectiles incident on lead. The
post-acceleration effect is quite large for low energies, tending to
disappear at $E/A\sim 1000$ and higher. The width of the momentum
distribution is not so sensitive to the energy, since it also carries an
important fraction which originates from the Fermi momentum of the ground
state wavefunction. Of course, our derivation depends on the assumptions
that (a) Coulomb dissociation is the major inelastic process, and that (b)
the fragments are focused at forward directions. However, these assumptions
are quite reasonable for systems with small binding energies, e.g., $^{11}Li$%
, and at intermediate energy collisions.

In conclusion, we have presented in this paper a simple analytical
derivation of the parallel velocity distribution of the fragments produced
in the elastic break-up of halo nuclei. We find a significant
postacceleration effect, in contrast to other works \cite{7,8,9}. Our
results are only valid if the breakup proceeds direct to the structureless
continuum, i.e. without a relevant resonance. The experimental data seems to
favor the first situation.

\bigskip

\noindent

{\large \bf Acknowledgments}

This work was partially supported by the FUJB/UFRJ, and by the
MCT/FINEP/CNPQ(PRONEX) under contract No. 41.96.0886.00.

\newpage
\noindent

{\large \bf Figure captions}

\vspace{3mm}

\begin{description}
\item[Figure 1]  {The parallel momentum distribution shift for $^{11}$Li
break-up on $^{208}$Pb as a function of the bombarding energy per nucleon.
Also shown is the width of the momentum distribution, }$\Gamma ,$ {\ as due
to the Coulomb dissociation process (solid line).}
\end{description}

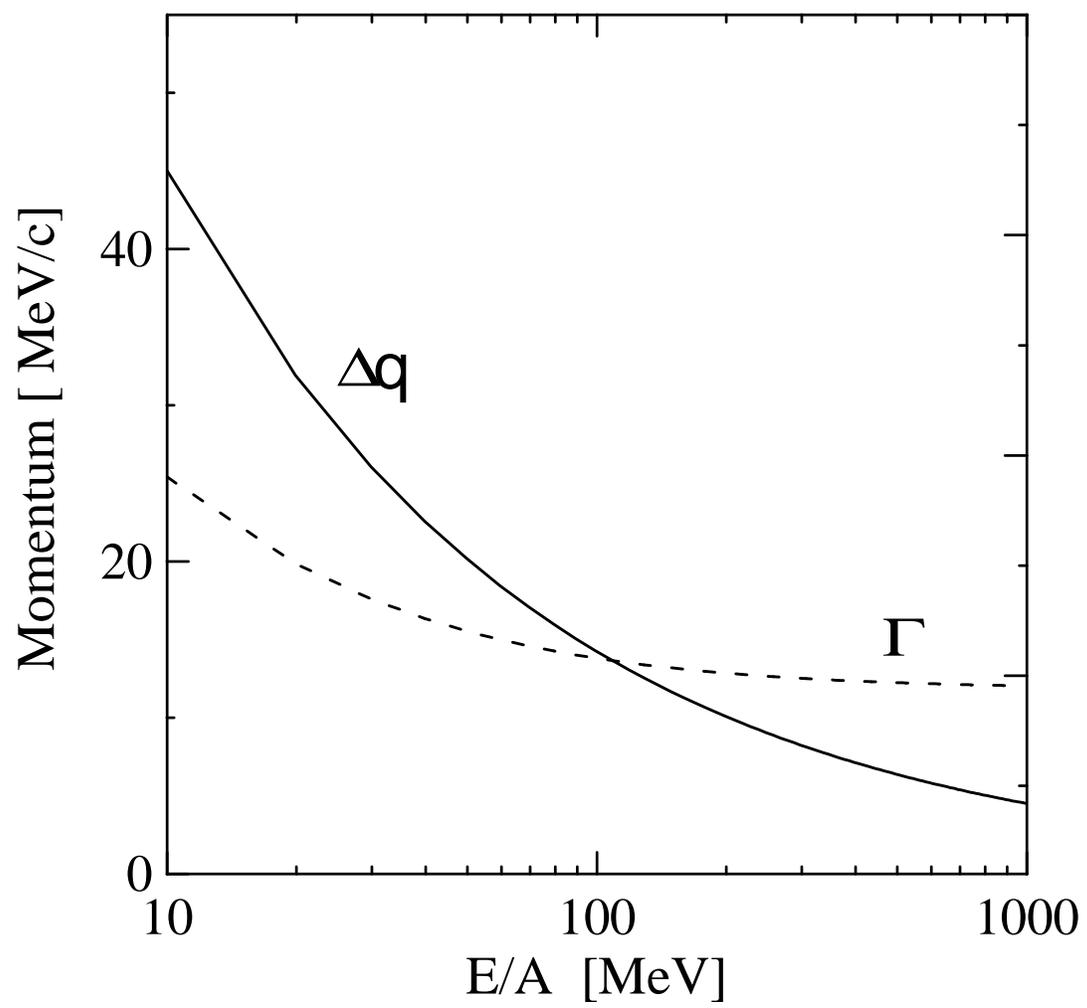
\end{document}